**Title:** Strong interactions and isospin symmetry breaking in a supermoiré lattice

**Authors**: Yonglong Xie[1,2,3]*[‡], Andrew T. Pierce[1]*[‡#], Jeong Min Park[2]*, Daniel E. Parker[1], Jie Wang[1,4], Patrick Ledwith[1], Zhuozhen Cai[1], Kenji Watanabe[5], Takashi Taniguchi[6], Eslam Khalaf[1], Ashvin Vishwanath[1], Pablo Jarillo-Herrero[2][‡], Amir Yacoby[1][‡]

**Affiliations:**
[1]*Department of Physics, Harvard University, Cambridge, MA 02138, USA*
[2]*Department of Physics, Massachusetts Institute of Technology, Cambridge, MA 02139, USA*
[3]*Department of Physics and Astronomy, Rice University, Houston, TX 77005, USA*
[4]*Center of Mathematical Sciences and Applications, Harvard University, Cambridge, MA 02138, USA*
[5]*Research Center for Electronic and Optical Materials, National Institute for Material Science, 1-1 Namiki, Tsukuba 305-0044, Japan*
[6]*Research Center for Materials Nanoarchitectonics, National Institute for Material Science, 1-1 Namiki, Tsukuba 305-0044, Japan*

*These authors contributed to this work.
[‡]Corresponding authors' emails: yx71@rice.edu, atp66@cornell.edu, pjarillo@mit.edu, yacoby@g.harvard.edu
[#]Present address: Kavli Institute at Cornell for Nanoscale Science, Ithaca, NY, USA

**Abstract:** In multilayer moiré heterostructures, the interference of multiple twist angles ubiquitously leads to tunable ultra-long-wavelength patterns known as supermoiré lattices. However, their impact on the system's many-body electronic phase diagram remains largely unexplored. We present local compressibility measurements revealing numerous incompressible states resulting from supermoiré-lattice-scale isospin symmetry breaking driven by strong interactions. By using the supermoiré lattice occupancy as a probe of isospin symmetry, we observe an unexpected doubling of the miniband filling near $\nu=-2$, possibly indicating a hidden phase transition or normal-state pairing proximal to the superconducting phase. Our work establishes supermoiré lattices as a tunable parameter for designing novel quantum phases and an effective tool for unraveling correlated phenomena in moiré materials.

**One-Sentence Summary**: Local compressibility measurements reveal ubiquitous supermoiré-scale symmetry breaking in multilayer moiré heterostructures.

**Main text**

The interplay between different length scales in low-dimensional electronic systems can radically alter their properties[1], providing a promising avenue for designing and discovering novel emergent phases. This principle has been demonstrated over the last decade by experiments on two-dimensional moiré superlattices in a magnetic field, beginning with the observation of fractal Hofstadter energy spectra[2–4] and exotic correlated topological states[5] in graphene aligned with hexagonal boron nitride (hBN), and later in magic-angle twisted bilayer graphene[6–18]. Systems with three or more layers provide new opportunities to further explore this design concept without requiring an external magnetic field, but by instead utilizing the interference between multiple moiré patterns. Such interference can lead to the emergence of a new superlattice distinct from the original moirés, which has been suggested as a key parameter for the observation of the quantum anomalous Hall effect in hBN aligned-twisted bilayer graphene[19]. A recent example that realizes this concept is twisted trilayer graphene (TTG), a system consisting of three consecutively stacked layers of graphene[20–22] (Fig. 1A) with small twist angles $\theta_{ij}$. Each rotation $\theta_{ij}$ is expected to produce a moiré pattern with wavelength $\lambda_{ij} \sim a/\theta_{ij}$, where $a$=0.246 nm is the graphene lattice constant. Notably, when $\theta_{12} \approx -\theta_{23}$ and thus $\theta_{13} \ll \theta_{12}, \theta_{23}$, the structure of TTG is well approximated by a complex double-periodic pattern characterized by two competing length scales (Fig. 1B). Indeed, scanning tunneling microscopy[23] and scanning transmission electron microscopy[24] experiments on TTG have revealed a long-wavelength periodic potential at the "moiré of moiré" wavelength $\lambda_{SM} \sim a/\theta_{13}$ —referred to below as a supermoiré lattice—that universally coexists with the parent moiré lattice with wavelength $\lambda_M \sim a/\theta_{12}$. The supermoiré length scale is considerably larger than that of the underlying moiré scale which prompts several key questions: Are interactions important on this length scale? Are the reported correlated phases in TTG, in particular superconductivity, related to the presence of a supermoiré lattice? And does the supermoiré lattice offer new possibilities for realizing new exotic quantum phases of matter?

The supermoiré lattice can modify the electronic properties of TTG in several ways. At the single-particle level, the supermoiré potential is anticipated to fold each of the parent moiré bands into a set of minibands that possess reduced bandwidth and rearranged quantum geometry as compared with the parent moiré bands[19,25–28] (Fig. 1C). The resulting minibands likely vary in the degrees to which they are perturbed by the supermoiré potential and to which their spectral weights follow the parent moiré versus the supermoiré lattice. These considerations suggest that some electrons may localize on the supermoiré lattice and the addition of the supermoiré potential may play an important role in the system's many-body phenomena. An immediate question that follows is whether the isospin symmetry, which is broken on the moiré scale, is also broken on the supermoiré scale despite the smallness of $U_{SM} \sim e^2/l_{SM}$ compared to $U_M \sim e^2/l_M$. Furthermore, TTG supports correlated states that break the translation symmetry of the parent moiré superlattice resulting from strong electron-electron interactions[29,30]. However, the incommensurability of the supermoiré lattice explicitly disrupts the parent moiré translation symmetry, calling into question whether such states remain favorable. These unresolved questions underscore the wealth of unexplored physics brought forth by the supermoiré lattice, which, to date, remains largely untapped.

In this work, we observe new strongly correlated phases enabled by the presence of a supermoiré lattice in a state-of-the-art magic-angle twisted trilayer graphene (MATTG) device. Our local thermodynamic measurements with a scanning single-electron transistor microscope reveal that the supermoiré lattice strongly modifies the topological moiré bands and gives rise to an unprecedentedly large set of incompressible states resulting from electron localization and supermoiré-lattice-scale isospin symmetry breaking. We observe unexpectedly large gaps associated with isospin symmetry breaking at the supermoiré lattice scale, many of which are comparable to or even greater than those associated with the parent moiré lattice. Strikingly, the supermoiré-scale isospin symmetry breaking persists down to zero magnetic field. On the electron side, the supermoiré potential introduces a duplicate set of isospin phase transitions mirroring those of the parent moiré bands, which is well captured by a mean-field model. However, near ν=−2, where superconductivity is present, we observe an apparent doubling of the number of electrons populating the supermoiré bands. These findings provide fresh insight into the long-standing puzzle of electron-hole asymmetric transport properties, especially concerning superconductivity, in the twisted graphene system. More broadly, our work suggests that supermoiré potentials may be used both as a new tuning knob for designing novel emergent phases and as an effective probe for unraveling complex correlated phenomena in moiré materials.

Our hBN-encapsulated MATTG device is fabricated via the "cut-and-stack" technique and placed on a metal back gate (see Methods and Fig. S1). Transport measurements on the device reveal the typical well-established phenomenology of MATTG[21,22,31,29], namely pronounced superconductivity on the hole side near ν=−2 and a sequence of resistive states most prominent under electron doping. In addition, we observe an additional weak resistance peak near and on the hole-doped side of the charge neutrality, suggestive of the supermoiré lattice. Overall, the phenomenology as revealed by transport is consistent with the known properties of MATTG.

**Signature of supermoiré lattice and incompressibility**

Examining the local compressibility within the same spatial region of the device reveals a significantly richer phase diagram than suggested by transport. Fig. 1D depicts the magnetic-field-dependent inverse compressibility dµ/dn as measured by the SET at a fixed location between the two voltage contacts (see Fig. S1A). An enormous number of incompressible peaks, corresponding to ground states with a thermodynamic gap, are observed. Conventionally, these states can be classified according to the Streda formula $\nu = C\frac{\phi}{\phi_0} + s$, where $\nu$ is the electron density per unit cell, $\phi$ is the magnetic flux per unit cell, $\phi_0 = h/e$ is the magnetic flux quantum; $C$ and $s$ are quantum numbers that characterize the Chern number and the number of filled bands per unit cell relative to charge neutrality. In the presence of two superlattices, electrons can be attached to either lattice and the value of $s$—the intercept of the incompressible state along the $\nu$ axis—can be further separated into $\nu_m$ and $\nu_{sm}$, which correspond to the number of electrons added to the moiré and supermoiré unit cells, respectively. Because of the spin and valley degrees of freedom, $v_m$ and $v_{sm}$ are expected to span from −4 to 4. In Fig. 1E, we identify all thermodynamic gaps in the phase diagram and classify them into one of four groups according to the ordered triple $(C, v_m, v_{sm})$. A complete table of the observed states can be found in the SI. First, the incompressible states possessing integer $v_m$ with $v_{sm}=0$ (grey lines in Fig. 1E) are simple Chern insulators (ChI) or integer quantum Hall (IQH) and are hallmarks of TTG readily observed in transport measurements[21,22,29]. Second, features with finite $C$, fractional $v_m$ and $v_{sm}=0$ (orange lines in Fig. 1E) are classified as symmetry broken Chern insulators (SBCIs), likely arise from the spontaneous breaking of the moiré unit cell translation symmetry and therefore represent topological charge density waves. Finally, we observe two new classes of gapped states with $v_{sm} \neq 0$ that we term sm-IQH/ChI (integer $v_m$) and sm-SBCI (fractional $v_m$), which demonstrate that the presence of supermoiré lattice strongly alters the electronic properties of the system and are the focus of this work. We also observe an intricate pattern of negative compressibility near charge neutrality, which is beyond the scope of this paper and will be addressed in future work[30].

The most immediate signature of the supermoiré lattice is the weak satellite incompressible peak at $v=-0.3$, which develops into a Landau fan with a principal Chern number sequence of $C=-6, -2$ and $+2$ (Fig. 2A and D). The observed fourfold sequence confirms that $v=-0.3$ marks the full filling of the super-moiré unit cell with four electrons due to spin and valley degeneracy, and we therefore identify these incompressible states as having $v_{sm}=-4$. The size of the supermoiré unit cell exceeds that of the principal moiré by a factor of $(0.3/4)^{-1} \sim 13$ and the corresponding wavelength is estimated to be 32.6 nm. These values are comparable to those from direct imaging experiments of supermoiré lattices[23,24] and correspond to an angle misalignment between the top and bottom graphene sheets of 0.43°. To further corroborate our identification of this feature as coming from the supermoiré lattice, we image a region of the device in which one of the graphene flakes terminates, leaving only a region of twisted bilayer graphene that allows us to independently determine the angle of the three graphene sheets (see SI section 3). Thus, the appearance of a satellite Landau fan near the charge neutrality point shows that the supermoiré lattice gives rise to a new set of bands at the single-particle level.

As the magnetic field is increased near the charge neutrality point, in addition to a typical sequence of isospin-symmetry-broken IQH states, referred to below as quantum Hall ferromagnets (QHFMs), we detect a series of satellite incompressible peaks that are offset along the $v$ axis by $\delta v = \pm 0.075$ (see Fig. S2). This density offset precisely translates to $\delta v_{sm}=\pm 1$, suggesting that this sequence of sm-IQH states comes about by filling one isospin-symmetry-broken supermoiré miniband in addition to the parent QHFM to maintain the same isospin polarization. We observe similar phenomenology at high magnetic field near the full-moiré filling densities of $v_m=\pm 4$, where additional families of incompressible states offset by $\delta v=\pm 0.075$ appear (Fig. 1E). Taken together, these observations suggest a general tendency for the isospin symmetry of the sm-IQH/ChI states to follow that of the corresponding parent moiré ChIs near the fillings $v=0$ and $\pm 4$.

**Supermoiré isospin symmetry breaking at zero field**

Remarkably, near the principal integer moiré fillings $\nu=+2$ and $+3$, we observe robust supermoiré-enabled incompressible states down to zero magnetic field, suggesting that the supermoiré electrons undergo a cascade of isospin phase transitions mirroring that of the parent moiré insulators. Specifically, at $\nu=+2$, we detect a sharp incompressible peak at zero field and a Landau level (LL) sequence of $+2$, $+4$, and $+6$. These observations are consistent with the expectation that strong interactions select two out of the four moiré bands to fill and form a correlated insulator. Strikingly, at the nearby filling $\nu=+2.15$, we observe an additional incompressible peak comparable in size to that at $\nu=2$, along with a "duplicated" twofold sequence of LLs. From the identical LL sequences at $\nu=2$ and $\nu=2.15$ we conclude that the sm-ChI at $\nu=2.15$ shares the same isospin polarization as the parent moiré $\nu=2$ insulator. Moreover, the density offset precisely matches that expected for two electrons per supermoiré unit cell, and further demonstrates the breaking of isospin symmetry on the supermoiré scale, which likely results from minimizing the exchange energy between electrons in moiré and supermoiré lattices. Turning our attention to $\nu=+3$, we observe a weaker zero-field incompressible peak together with a nondegenerate LL sequence of $+2$, $+3$, $+4$, corresponding to fully broken isospin symmetry. Like the behavior at $\nu=2$, a satellite incompressible sm-ChI peak with comparable magnitude is also present at the nearby filling $\nu=+3.075$, with the offset $\delta\nu = +0.075$ reflecting the single remaining supermoiré band. Thus, the isospin symmetry of the sm-ChIs about $\nu=3$ follows that of the parent ChIs in a manner resembling $\nu=2$.

The salient features of our data at low magnetic field are surprisingly well captured by a phenomenological mean-field model with strong interactions between all electrons, both moiré and supermoiré. Due to the strong separation of moiré and supermoiré length scales, we anticipate that the formation of large regions of alternating-twist mirror symmetric trilayer separated by mirror-symmetry-breaking domain walls, as shown in recent direct imaging experiments[23,24]. Motivated by these considerations, we assume that the electrons in the system can be separated into two species according to whether they inhabit the moiré or supermoiré lattice and we consider a phenomenological density of states for each species, with smaller bandwidth for supermoiré electrons due to their larger length scale. We include strong contact interactions $U_{m-m}$, $U_{m-sm}$, and $U_{sm-sm}$, that are independent of spin and valley. The model may then be solved (see SI for full details) at mean-field level under standard approximations[32,14,16], yielding the occupations of each species and the compressibility as a function of filling shown in Figs 2J-O. Near charge neutrality, where the effects of interaction-induced dispersion are expected to be weakest due to the vanishing Hartree potential, our model reproduces the satellite peak observed at $\nu=-0.3$, validating the choice of the single-particle parameters in our model. Near $\nu=+2$ and $\nu=+3$, where Hartree effects are significant, the computed compressibility curves closely resemble those observed experimentally: in both cases, only one satellite peak appears at higher density than the parent moiré ChI. This occurs because Stoner ferromagnetism ensures that an equal number of moiré and supermoiré isospin species are populated together at integer fillings. Additional doping then populates the supermoiré electrons due to their lower kinetic energy until they are fully filled, leading to a satellite incompressible state. We note these results are robust, depending only weakly on the density of states profile, as long as the kinetic energy of supermoiré electrons is substantially smaller than that of moiré electrons (see Fig. S5). Overall, the excellent agreement between our data and model calculations strongly indicates that the Coulomb interactions involving the supermoiré electrons are also strong, making them key to understanding the correlated phases in the system.

**Isospin symmetry anomaly near v=−2**

The behavior of the isospin degree of freedom on the hole-doped side stands in marked contrast to that on the electron-doped side and cannot be explained with the phenomenological model described above. At $v=-2$, we observe a sharp incompressible peak at zero magnetic field and a LL sequence of −2, −4, −6, ..., consistent with twofold degeneracy. While our model predicts a supermoiré isospin transition near $v=-2.15$, we instead detect a broad peak near $v=-2.3$. This unexpected doubling of the supermoiré filling is confirmed by the presence of LLs emanating from $v=-2.3$: we find a LL sequence of −2, −4, −8, which, after accounting for the $C=-2$ offset from the $N=0$ LL of the Dirac bands, translates to 0, −2, −6, identical to that observed at $v=-0.3$. Incidentally, the recovery of fourfold supermoiré filling coincides with the density range at which superconductivity is observed in transport measurements. By contrast, superconductivity is absent on the electron-doped side where isospin symmetry is broken.

A natural interpretation of the apparent doubling of supermoiré filling is a hidden phase transition between $v=-2$ and $v=-2.3$ from an isospin symmetry-broken phase to a fourfold isospin-symmetric state. Such a phase transition may generally be present in TTG but may only be detectable by examining its interplay with the supermoiré potential using local imaging techniques like our high-resolution local compressibility measurements.

Alternatively, the doubling of supermoiré filling can result from the formation of paired charge $2e$ objects rather than individual electrons, which will produce the feature at $v=-2.3$ while maintaining the broken two-fold isospin degeneracy. We note that this interpretation is consistent with the pseudogap-like behavior observed in STM studies[33,34] between $v=-2$ and $-3$ at energy scales considerably greater than $T_c$, which can be interpreted as a signature for pre-formed pairs. The supermoiré potential is likely to favor the formation of commensurate charge density wave orders of these charge $2e$ objects leading to an incompressible feature at $v=-2.3$. Further theoretical and experimental efforts are required to identify the microscopic mechanism for the origin of the observed doubling of supermoiré filling and its relation to the superconductivity in MATTG.

To firmly correlate the isospin symmetry characteristics inferred from the local compressibility measurements with the presence of the supermoiré lattice, we perform spatially resolved compressibility measurement over a 6 μm span. We first present measurements taken at $B=2$ T (Fig. 4A-B), where the absence of the Dirac bands other than their $N=0$ LLs enables us to estimate the isospin degeneracy at various fillings most accurately. Strikingly, features associated with supermoiré potential (Fig. 4B, blue) are visible over the entire range, despite the variations in the exact density at which they emerge. At each location, we quantify the local twist angle and supermoiré wavelength (Fig. 4C-D) by comparing the position of the full filling IQH/ChI (±2, ±4) and the separation between the (−6, 0 −4) and (−6, 0, 0), respectively. Over 6 μm, the local twist angle varies by 0.03° and the supermoiré wavelength varies by 10 nm, corresponding to angle mismatches from 0.4° to 0.48°. The variations in local twist angle are not correlated with those in the mismatch. Next, using the same data set, we characterize the local filling dependent isospin degeneracy by extracting the supermoiré density offset in units of electrons per supermoiré unit cell (see SI for full details). We find that, at all positions investigated, the system retains an isospin degeneracy of four at $\nu=-2$, while at filling $\nu=+2$ the degeneracy is reduced to two and at filling $\nu=+3$ the degeneracy is reduced to one. Importantly, at zero magnetic field, all the principal moiré and supermoiré incompressible peaks analyzed above are observed to be robust over the full 6 μm range (Fig. S3), suggesting that the pattern of symmetry breaking identified in Figs. 2 and 3 holds globally to zero magnetic field.

**Strong interactions and topology**

The dominant role of supermoiré potential in generating novel quantum phases is made even more evident by the observation of several unexpected incompressible states. First, the sm-ChIs and sm-SBCIs may exhibit different order from that of neighboring moiré states. For example, the parent moiré ChI with (+5, +1, 0) corresponds to a spin and valley polarized Chern insulator (Fig. 5A and 5D) and the sm-ChI with (+5, +1, −1) likely prefers to maintain the same isospin polarization by removing one hole from a $C=0$ supermoiré band, similar to the satellite peaks near the principle QHFM states. By contrast, the (+5, +1, −2) sm-ChI requires the removal of two holes relative to the (+5, +1, 0) parent state, likely indicating that the sm-ChI possesses a different isospin structure from that of the parent state. Second, the thermodynamic gaps of the sm-ChIs and sm-SBCIs can be comparable to or greater than neighboring moiré states, calling into question which ground states should be viewed as the parent states. For example, although the (+5, +1/2, +1) sm-SBCI occurs near a moiré SBCI with (+5, +1/2, 0), the sm-SBCI state exhibits a considerably larger gap and appears over a much larger magnetic field range than the neighboring moiré SBCI (Fig. 5B and 5E). These characteristics suggest that sm-ChIs and sm-SBCIs may originate from a distinct mechanism that does not rely on the existence of a parent moiré insulator. This picture is further supported by the observation of the sm-SBCI state (+3, +3/2, +1), which lacks an immediate parent moiré incompressible state in its proximity (Fig. 5C and 5F).

Finally, our high-resolution local compressibility measurements reveal the rich interplay between the supermoiré potential and the topological moiré bands as a function of carrier density. First, near $\nu=0$, we observe a sequence of sm-IQH states that duplicates the sequence of Chern numbers of the principal IQH states (Fig. S6A and S6C). The same behavior is observed for the IQH and sm-IQH sequences at $\nu=\pm4$ (Fig. S6B and S6D), as well as at $\nu=+2$ (Fig. 2B and 2E, where the sequence of sm-ChIs emanating from $\nu=2.15$ exactly duplicates the Chern numbers of the parent ChIs. Since the sm-IQH and sm-ChIs share the same Chern number as the parent IQH and ChI states, we conclude that the filled supermoiré minibands near fillings $v=0$, $+2$ and $\pm 4$ contribute a total Chern number of zero. Unexpectedly, however, the LL sequences emanating from $v=+3$ and $v=3.075$ are different (Fig. 2C and 2F), with the most prominent difference being the increment of 1 in the total Chern number between $(+2, +3, 0)$ versus $(+3, +3, +1)$. The parent moiré state $(+2, +3, 0)$ is analogous to the $C=0$ state at $v=+3$ in twisted bilayer graphene which may be identified as a translation symmetry breaking state constructed by occupying all parent $C=\pm 1$ bands with only a set of $C=0$ bands remaining empty[16]. However, in this scenario, the remaining supermoiré minibands must possess $C=0$ and cannot produce the observed valley-polarized $(+3, +3, +1)$ state. One possible resolution of this problem is that approximate moiré translation symmetry is recovered via a phase transition between the $(+2, +3, 0)$ and $(+3, +3, +1)$ states. Then, the $(+3, +3, +1)$ state is naturally obtained by filling one $C=0$ miniband relative to a valley-polarized $(+3, +3, 0)$ insulator. The recovery of translation symmetry and valley polarization may result from the incommensurability of the supermoiré lattice disrupting the parent moiré translation symmetry and simultaneously weakening the Hartree potential responsible for translation symmetry breaking as electrons occupy supermoiré lattice sites. The possible existence of such a phase transition is consistent with the strong negative compressibility that separates these two states. Alternatively, the supermoiré miniband that is filled between $(+2, +3, 0)$ and $(+3, +3, +1)$ may itself carry $C=+1$. Further theoretical studies and experiments are required to determine the topology of the supermoiré minibands from the observed sequence of incompressible states. Regardless of the specific nature of the ground states, we emphasize that these Chern number sequences hold over the full 6 μm range investigated (Figs. 4A-B), indicating that the apparent reconstruction of the ground state band topology at $v=+3$ as probed by the supermoiré occurs generally.

Taken together, the abundance of incompressible states observed in our experiment suggests that the supermoiré potential may be used as a novel tuning knob for designing correlated quantum phases in twisted multilayers. The observation of isospin symmetry-broken incompressible states on the supermoiré lattice highlights its usefulness as a novel probe of isospin symmetry in moiré systems. The contrasting isospin symmetry patterns on the electron-doped side versus that on the hole-doped side provides an important constraint for understanding the superconductivity in TTG and may be indispensable in describing superconducting quadri- and pentalayer graphene[35–37] as the effect of the supermoiré potential is expected to become more important as the number of graphene layers is increased.

## References

1. Yankowitz, M. *et al.* Emergence of superlattice Dirac points in graphene on hexagonal boron nitride. *Nature Phys* **8**, 382–386 (2012).
2. Hunt, B. *et al.* Massive Dirac Fermions and Hofstadter Butterfly in a van der Waals Heterostructure. *Science* **340**, 1427–1430 (2013).
3. Dean, C. R. *et al.* Hofstadter's butterfly and the fractal quantum Hall effect in moiré superlattices. *Nature* **497**, 598–602 (2013).
4. Ponomarenko, L. A. *et al.* Cloning of Dirac fermions in graphene superlattices. *Nature* **497**, 594–597 (2013).
5. Spanton, E. M. *et al.* Observation of fractional Chern insulators in a van der Waals heterostructure. *Science* **360**, 62–66 (2018).
6. Cao, Y. *et al.* Superlattice-Induced Insulating States and Valley-Protected Orbits in Twisted Bilayer Graphene. *Phys. Rev. Lett.* **117**, 116804 (2016).
7. Cao, Y. *et al.* Correlated insulator behaviour at half-filling in magic-angle graphene superlattices. *Nature* **556**, 80–84 (2018).
8. Cao, Y. *et al.* Unconventional superconductivity in magic-angle graphene superlattices. *Nature* **556**, 43–50 (2018).
9. Tomarken, S. L. *et al.* Electronic Compressibility of Magic-Angle Graphene Superlattices. *Phys. Rev. Lett.* **123**, 046601 (2019).
10. Wu, S., Zhang, Z., Watanabe, K., Taniguchi, T. & Andrei, E. Y. Chern insulators, van Hove singularities and topological flat bands in magic-angle twisted bilayer graphene. *Nature Materials* **20**, 488–494 (2021).
11. Nuckolls, K. P. *et al.* Strongly correlated Chern insulators in magic-angle twisted bilayer graphene. *Nature* **588**, 610–615 (2020).
12. Saito, Y. *et al.* Hofstadter subband ferromagnetism and symmetry-broken Chern insulators in twisted bilayer graphene. *Nature Physics* **17**, 478–481 (2021).
13. Das, I. *et al.* Symmetry-broken Chern insulators and Rashba-like Landau-level crossings in magic-angle bilayer graphene. *Nature Physics* 1–5 (2021) doi:10.1038/s41567-021-01186-3.
14. Park, J. M., Cao, Y., Watanabe, K., Taniguchi, T. & Jarillo-Herrero, P. Flavour Hund's coupling, Chern gaps and charge diffusivity in moiré graphene. *Nature* **592**, 43–48 (2021).
15. Choi, Y. *et al.* Correlation-driven topological phases in magic-angle twisted bilayer graphene. *Nature* **589**, 536–541 (2021).
16. Pierce, A. T. *et al.* Unconventional sequence of correlated Chern insulators in magic-angle twisted bilayer graphene. *Nat. Phys.* **17**, 1210–1215 (2021).
17. Yu, J. *et al.* Correlated Hofstadter spectrum and flavour phase diagram in magic-angle twisted bilayer graphene. *Nat. Phys.* **18**, 825–831 (2022).
18. Xie, Y. *et al.* Fractional Chern insulators in magic-angle twisted bilayer graphene. *Nature* **600**, 439–443 (2021).

19. Shi, J., Zhu, J. & MacDonald, A. H. Moiré commensurability and the quantum anomalous Hall effect in twisted bilayer graphene on hexagonal boron nitride. *Phys. Rev. B* **103**, 075122 (2021).
20. Khalaf, E., Kruchkov, A. J., Tarnopolsky, G. & Vishwanath, A. Magic angle hierarchy in twisted graphene multilayers. *Phys. Rev. B* **100**, 085109 (2019).
21. Park, J. M., Cao, Y., Watanabe, K., Taniguchi, T. & Jarillo-Herrero, P. Tunable strongly coupled superconductivity in magic-angle twisted trilayer graphene. *Nature* **590**, 249–255 (2021).
22. Hao, Z. *et al.* Electric field–tunable superconductivity in alternating-twist magic-angle trilayer graphene. *Science* **371**, 1133–1138 (2021).
23. Turkel, S. *et al.* Orderly disorder in magic-angle twisted trilayer graphene. *Science* **376**, 193–199 (2022).
24. Craig, I. M. *et al.* Local atomic stacking and symmetry in twisted graphene trilayers. Preprint at http://arxiv.org/abs/2303.09662 (2023).
25. Anđelković, M., Milovanović, S. P., Covaci, L. & Peeters, F. M. Double Moiré with a Twist: Supermoiré in Encapsulated Graphene. *Nano Lett.* **20**, 979–988 (2020).
26. Leconte, N. & Jung, J. Commensurate and incommensurate double moire interference in graphene encapsulated by hexagonal boron nitride. *2D Mater.* **7**, 031005 (2020).
27. Zhu, Z., Cazeaux, P., Luskin, M. & Kaxiras, E. Modeling mechanical relaxation in incommensurate trilayer van der Waals heterostructures. *Phys. Rev. B* **101**, 224107 (2020).
28. Zhu, Z., Carr, S., Massatt, D., Luskin, M. & Kaxiras, E. Twisted Trilayer Graphene: A Precisely Tunable Platform for Correlated Electrons. *Phys. Rev. Lett.* **125**, 116404 (2020).
29. Shen, C. *et al.* Dirac spectroscopy of strongly correlated phases in twisted trilayer graphene. *Nat. Mater.* **22**, 316–321 (2023).
30. Pierce, A. T. *et al.* Interplay between light and heavy electrons in twisted trilayer graphene. *(in preparation)*.
31. Liu, X., Zhang, N. J., Watanabe, K., Taniguchi, T. & Li, J. I. A. Isospin order in superconducting magic-angle twisted trilayer graphene. *Nat. Phys.* **18**, 522–527 (2022).
32. Zondiner, U. *et al.* Cascade of phase transitions and Dirac revivals in magic-angle graphene. *Nature* **582**, 203–208 (2020).
33. Oh, M. *et al.* Evidence for unconventional superconductivity in twisted bilayer graphene. *Nature* **600**, 240–245 (2021).
34. Kim, H. *et al.* Evidence for unconventional superconductivity in twisted trilayer graphene. *Nature* **606**, 494–500 (2022).
35. Park, J. M. *et al.* Robust superconductivity in magic-angle multilayer graphene family. *Nat. Mater.* **21**, 877–883 (2022).
36. Burg, G. W. *et al.* Emergence of correlations in alternating twist quadrilayer graphene. *Nat. Mater.* **21**, 884–889 (2022).
37. Zhang, Y. *et al.* Promotion of superconductivity in magic-angle graphene multilayers. *Science* **377**, 1538–1543 (2022).

**Acknowledgments:** We acknowledge discussions with Ady Stern and Aviram Uri. **Funding**: This work was sponsored by the Army Research Office under award number W911NF-21-2-0147 and by the Gordon and Betty Moore Foundation EPiQS initiative through Grant GBMF 9468. Help with transport measurements and data analysis were supported by the National Science Foundation (DMR-1809802), and the STC Center for Integrated Quantum Materials (NSF Grant No. DMR-1231319). P.J-H also acknowledges support from the Gordon and Betty Moore Foundation's EPiQS Initiative through Grant GBMF 9463. A.T.P. and P.J.L. acknowledge support from the Department of Defense through the National Defense Science and Engineering Graduate Fellowship (NDSEG) Program. Y.X. acknowledges partial support from the Harvard Quantum Initiative in Science and Engineering. Y.X, A.T.P. and A.Y. acknowledge support from the Harvard Quantum Initiative Seed Fund. A.V. was supported by a Simons Investigator award and by the Simons Collaboration on Ultra-Quantum Matter, which is a grant from the Simons Foundation (651440, AV). E.K. was supported by a Simons Investigator Fellowship, by NSF-DMR 1411343, and by the German National Academy of Sciences Leopoldina through grant LPDS 2018-02 Leopoldina fellowship. This research is funded in part by the Gordon and Betty Moore Foundation's EPiQS Initiative, Grant GBMF8683 to D.E.P. K.W. and T.T. acknowledge support from the JSPS KAKENHI (Grant Numbers 21H05233 and 23H02052) and World Premier International Research Center Initiative (WPI), MEXT, Japan. This work was performed, in part, at the Center for Nanoscale Systems (CNS), a member of the National Nanotechnology Infrastructure Network, which is supported by the NSF under award no. ECS-0335765. CNS is part of Harvard University. **Author Contributions**: Y.X., A.T.P., J.M.P., P.J.-H. and A.Y. conceived and designed the experiment. A.T.P., Y.X. and Z.C. performed the scanning SET experiment and analyzed the data with input from A.Y. J.M.P. and P.J.-H. designed and provided the samples and contributed to the analysis of the results. D.E.P., J.W., P.J.L. E.K. and A.V. performed the theoretical analysis. K.W. and T.T. provided hBN crystals. All authors participated in discussions and in the writing of the manuscript.

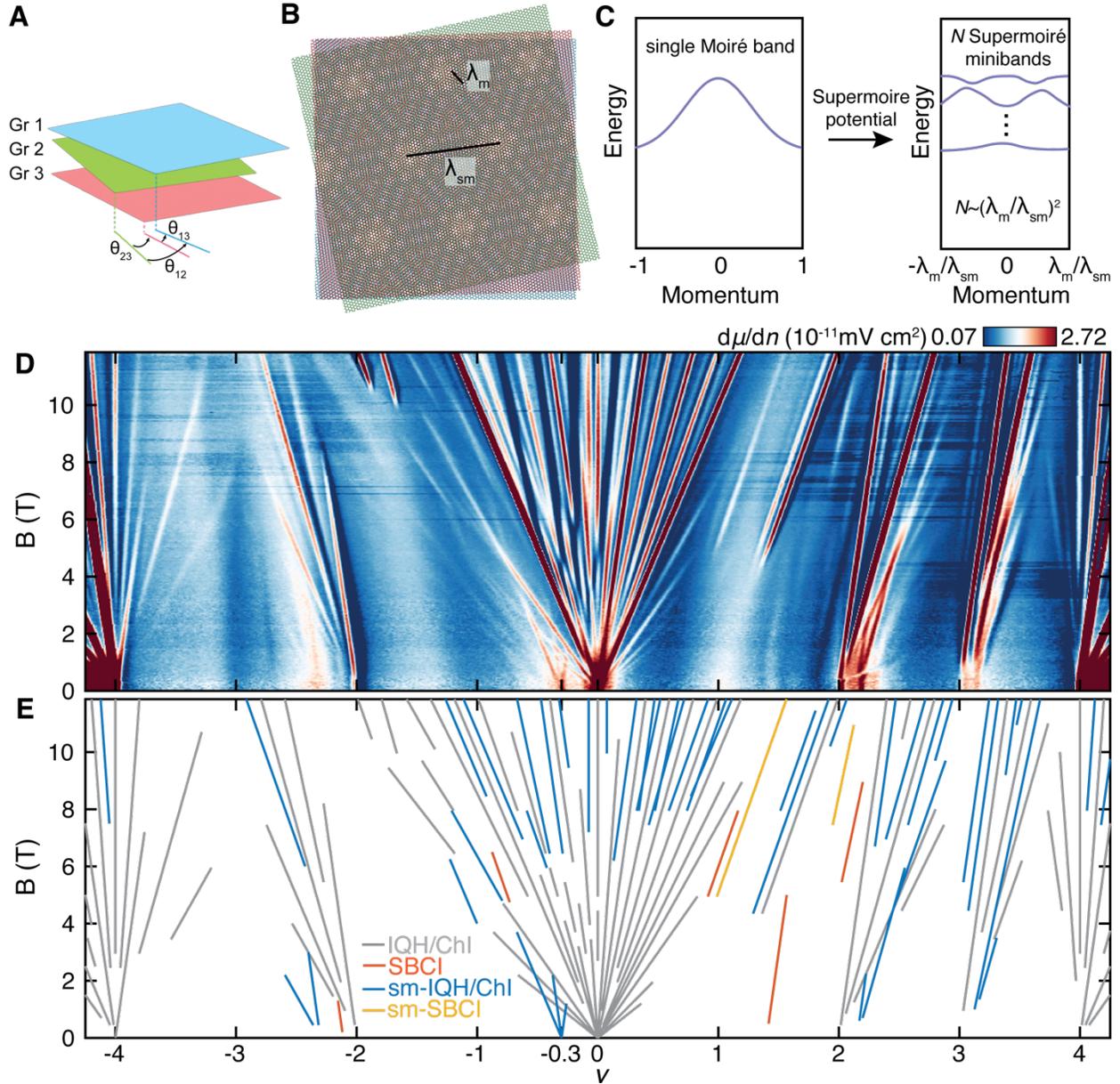

**Fig. 1. Incompressible states at commensurate and incommensurate fillings of the moiré unit cell.** (**A**) TTG formed by consecutively stacking three layers of graphene (labeled Gr 1-3) small twist relative angles $\theta_{ij}$. (**B**) Sketch of TTG lattice exhibiting a principal moiré wavelength $\lambda_m$ and a larger supermoiré wavelength $\lambda_{sm}$. (**C**) Sketch of a single parent moiré flat band in momentum space (left panel) and the sequence of minibands that results after Brillouin zone folding (right panel). (**D**) Local inverse compressibility d$\mu$/d$n$ plotted as a function of electron density, in units of electrons per moiré unit cell ν, and magnetic field *B*. (**E**) Wannier diagram identifying the incompressible states present in (A). Grey and orange lines correspond to integer quantum Hall states or Chern insulators and symmetry broken Chern insulators emanating from integer or a commensurate fraction of the moiré unit cell of twisted trilayer graphene. Blue and yellow lines correspond to IQH/ChI and SBCI with a non-zero supermoiré band filling.

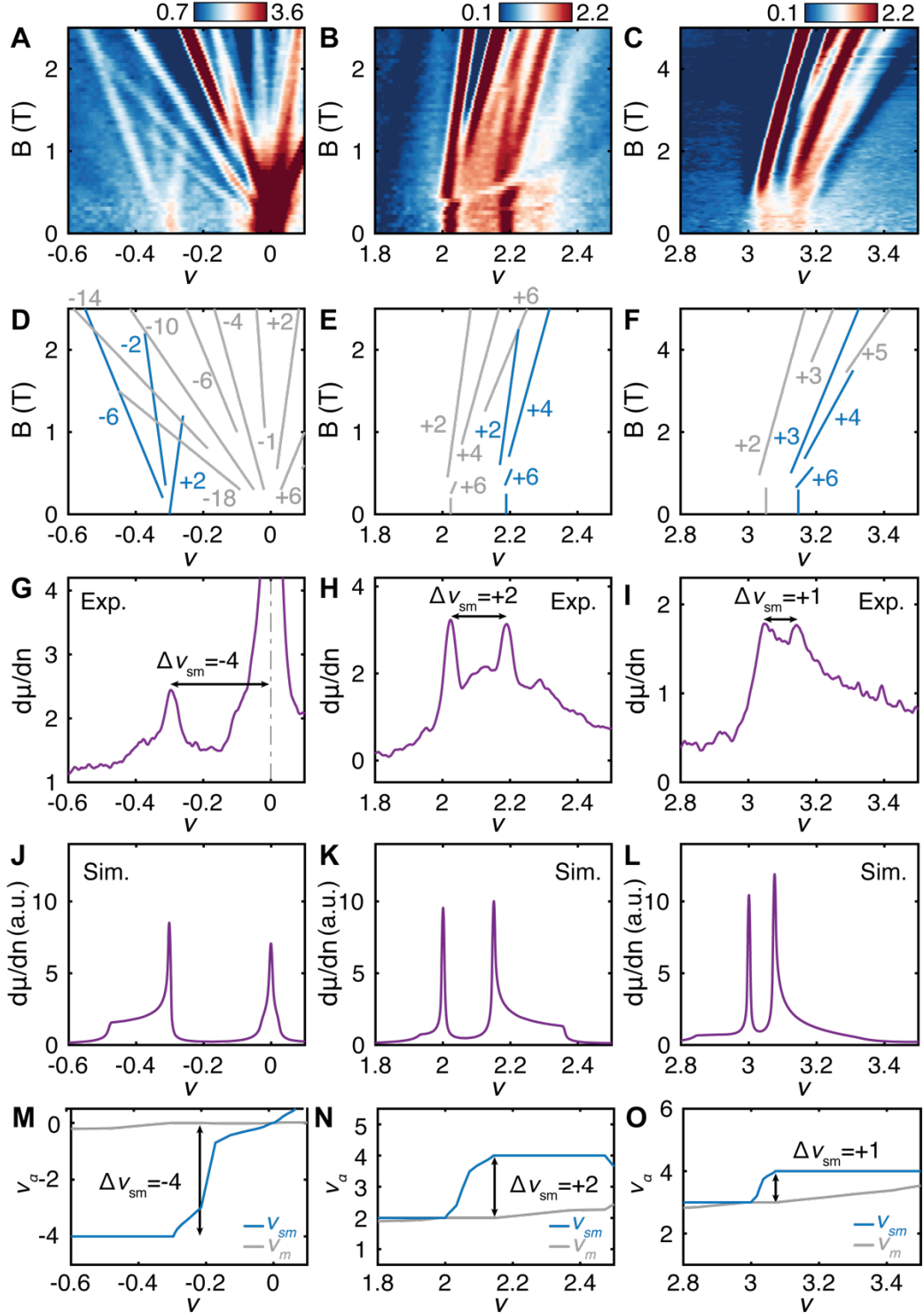

Fig. 2. Supermoiré isospin symmetry breaking down to zero magnetic field. (A-D) Zoomed-in view of local inverse compressibility near ν=0 (A), ν=+2 (B) and ν=+3 (C). (D-F) Wannier diagrams highlighting the incompressible states observed in A-C colored according to the classification of Fig. 1B. Grey lines denote states emanating from integer fillings of the moiré unit cell and blue lines correspond to states where an integer number of additional supermoiré

minibands are filled or emptied. (**G-I**) Experimental dµ/dn line traces near near ν=0 (G), ν=+2 (H) and ν=+3 (I) and averaged over 0 to 200 mT. The density offsets between the incompressible peaks approximately correspond to −4, +2, and +1 electron(s) per supermoiré unit cell respectively. (**J-L**) simulated dµ/dn traces near ν=0 (J), ν=+2 (K) and ν=+3 (**L**) calculated according to the mean-field model exhibit density offsets matching those in G-I. (**M-O**) simulation results for the total supermoiré filling near ν=0 (M), ν=+2 (N) and ν=+3 (O).

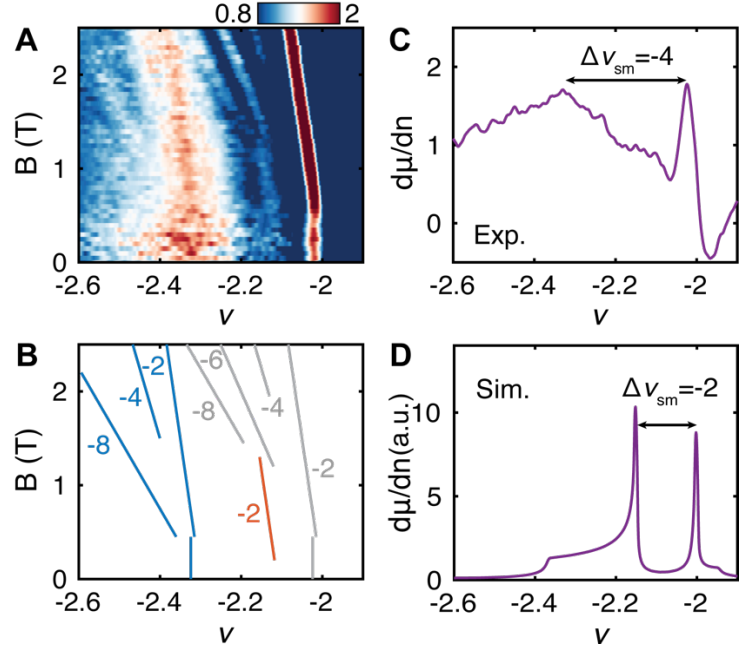

**Fig. 3. Isospin symmetry anomaly near $v=-2$ probed by the supermoiré lattice. (A)** Zoomed-in view of local inverse compressibility $d\mu/dn$ ($\times 10^{-11}$ mV cm$^{-2}$) near $v=-2$. **(B)** Wannier diagram highlighting the incompressible states observed in (A) colored according to the classification of Fig. 1B. **(C)** Experimental $d\mu/dn$ line trace near $v=-2$ averaged over 0 to 200 mT. The density offset between the broad incompressible peak and the moiré correlated insulator corresponds to a filling of four additional holes per supermoiré unit cell in agreement with the extrapolated intercepts of the LL sequences observed to emerge starting around 1 T. **(D)** Mean-field simulation results near $v=-2$ predict the appearance of an incompressible state at a filling of two holes per supermoiré unit cell in disagreement with the experiment, suggesting the presence of beyond-mean-field electronic correlations.

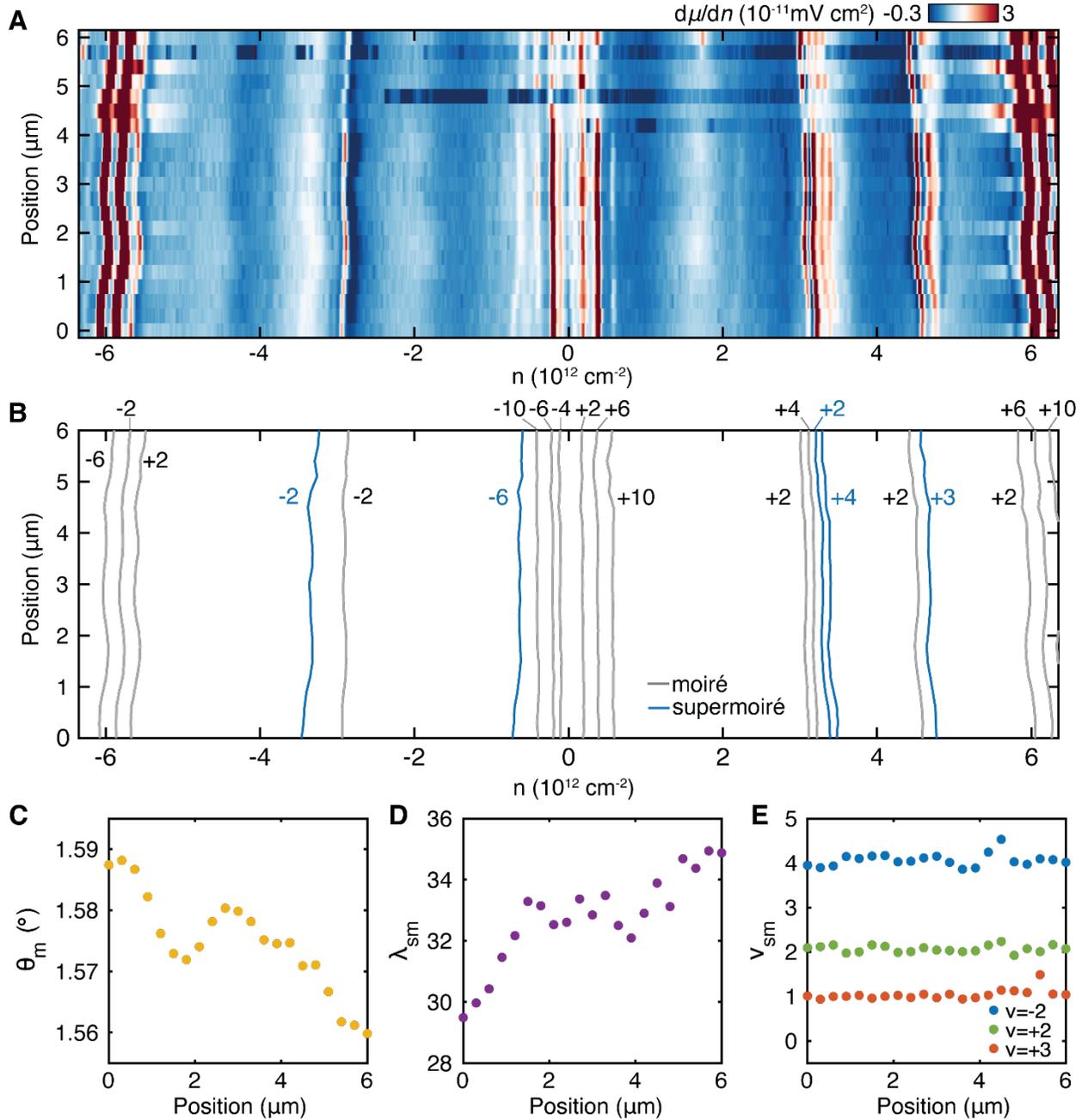

**Fig. 4. Ubiquitous supermoiré isospin order.** (**A**) Spatial dependence of local inverse compressibility dμ/d$n$ as a function of electron density measured at $B$=2 T. (**B**) Trajectories of the incompressible states observed in (A). The numbers denote the Chern number of each incompressible state. (**C-D**) Local twist angle (C) and supermoiré wavelength (D) estimated using the data in (A) (see Methods). (**E**) Isospin flavor degeneracy near moiré filling of −2, +2 and +3. as a function of distance.

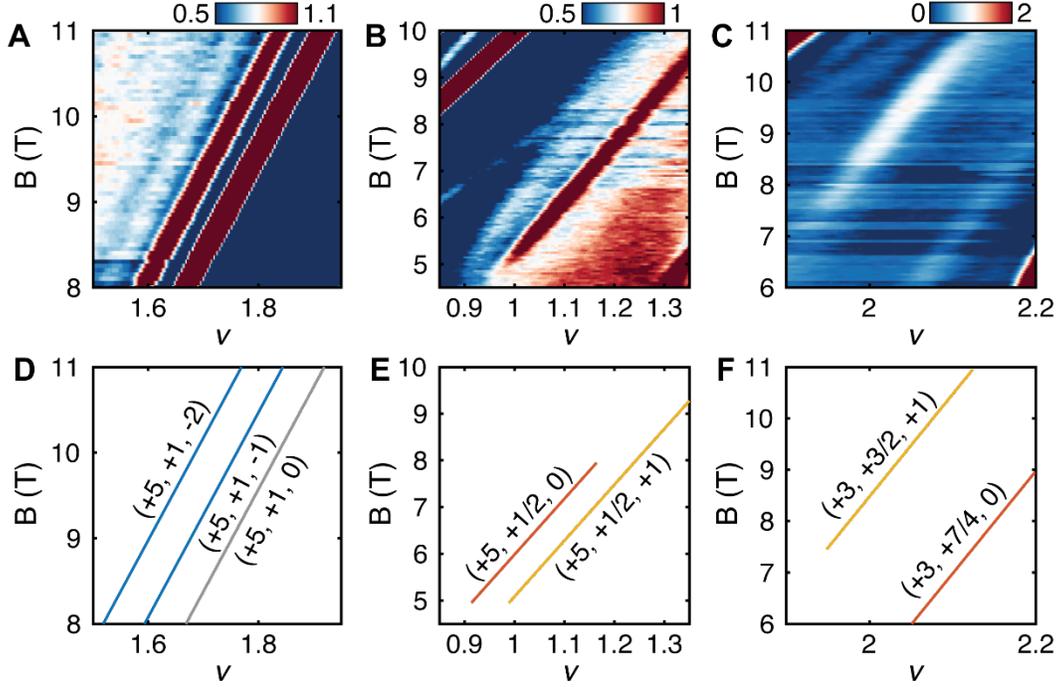

**Fig. 5. Additional supermoiré states. (A-C)** Measurements of d$\mu$/d$n$ (×10$^{-11}$ mV cm$^{-2}$) in various density ranges between 6 and 12 T showing additional supermoiré states. **(D-F)** Wannier diagrams corresponding to the states observed in (A-C) colored according to the classification used in Fig. 1B. Grey, orange, blue, yellow lines denote the ChIs, SBCI, sm-ChIs and sm-SBCIs, respectively.